\documentclass[11pt]{article}
\usepackage{amsmath,amssymb,amsfonts}
\usepackage[dvips]{graphicx}
\usepackage{epsfig}
\usepackage{cite}

\def\e{\epsilon}

\def\i{\iota}

\def\s{\sigma}

\def\p{\partial}

\begin{document}
\begin{titlepage}

\vfill
\begin{center}
{\Large\bf Quantum Mechanical Hamiltonians \\ with Large Ground-State Degeneracy}

\vskip 1cm

Choonkyu Lee{\let\thefootnote\relax\footnote{$^\dagger$
Electronic address : cklee@phya.snu.ac.kr}}$^\dagger$

{\it Department of Physics and Astronomy and Center for Theoretical Physics
\\Seoul National University, Seoul 151-147, Korea}
\vskip 2mm
and 
\vskip 2mm
Kimyeong Lee{\let\thefootnote\relax\footnote{$^\ddagger$ 
Electronic address : klee@kias.re.kr}}$^\ddagger$

{\it Korea Institute
for Advanced Study, Seoul 130-722, Korea}

\end{center}
\vfill

\begin{abstract}
Nonrelativistic Hamiltonians  with    large,
even infinite, ground-state degeneracy   are studied by connecting the degeneracy to the property of a Dirac operator. We then
identify a special class of Hamiltonians, for which the full space of degenerate ground
states in any spatial dimension can be exhibited explicitly.
The two-dimensional version of the latter coincides with
the Pauli Hamiltonian, and recently-discussed models
leading to higher-dimensional Landau levels are obtained
as special cases of the higher-dimensional version of this
Hamiltonian. But, in our framework, it is only the asymptotic
behavior of the background `potential' that matters for the
ground-state degeneracy.  We work out in detail the ground states of the three-dimensional model in the presence of a uniform magnetic field and such   potential. In the latter case one can see degenerate stacking of all 2d Landau levels along the magnetic field axis. \end{abstract}

\vfill
\end{titlepage}

\parskip 0.1 cm
\newpage
\renewcommand{\thefootnote}{\#\arabic{footnote}}
\setcounter{footnote}{0}

\parskip 0.2 cm

\section{Introduction}
Quantum mechanical Hamiltonians admitting highly
degenerate   energy eigenstates are rather special, and
they can thus serve as  useful theoretical models  to
explain some novel properties exhibited by physical
systems under certain circumstances. The most notable
example is provided by the integer quantum Hall effect
in two-dimensional (2d) electron gas, for which the
quantized, and infinitely degenerate, energy eigenspace
structure of the 2d Landau Hamiltonian \cite{Landau}
is largely responsible. The study of the 2d quantum
Hall effect, including both integer and fractional ones,
has been greatly benefitted by the elegant analytic
properties of the Landau level wave functions (not only
in Euclidean space, but also in curved or topologically
different backgrounds \cite{Haldane}). In this 2d Landau
Hamiltonian, the vector potential describing the
background magnetic field is Abelian. 

Recently, it has
been noticed by various authors that there also exist
higher-dimensional, especially three-dimensional (3d),
Hamiltonians the eigenstate structure of which exhibit
Landau-level-like degeneracy
\cite{Zhang-Hu, Li-Wu, Li-Intriligator, Li-Zhang}.
A distinctive feature from the 3d Euclidean-space Landau systems   of Refs.
\cite{Li-Wu,Li-Intriligator,Li-Zhang}
is that the related Hamiltonian typically contains spin-orbit
coupling term ${\bf E}\times {\bf p} \cdot  {\boldsymbol \sigma}$
(for a radial electric field ${\bf E}$ for instance),   which can be
accommodated within the standard `magnetic Hamiltonian'
by having the spin-$\frac12$ particle couple to a
\emph{non-Abelian} SU(2) gauge potential of the form
$ {\cal A} = {\bf E} \times {\boldsymbol \sigma}$. One may
expect that some nontrivial physical applications
utilizing the latter Hamiltonians come out in the future.

It would  certainly be useful to know  something on
the Hamiltonian `form' that can give rise to such highly
degenerate eigenstate structure, especially for the
ground state of the system. In the case of a nonrelativistic
particle described by a spinor wave function in n-dimensional
Euclidean (configuration) space, we may take the Hamiltonian
to have the general form
\begin{equation}\label{H}
{\cal H} = \frac{1}{2M}\Big( {\bf p}\cdot{\bf p} + 
{\cal P} ({\bf x, S})\cdot{\bf p}   + {\cal Q({\bf x, S})}\Big)\ ,
\end{equation}
where   ${\bf{x}} = (x^1,\cdots x^n)$ denote position
coordinates, ${\bf p}=-i\hbar \nabla$   the momentum differential operators, and
${\cal S} = \{S_{ij}; 1\leq i <j\leq n\}$ represent
related spin generators. 
(If the Hamiltonian involves functions ${\cal P}_i({\bf x, S})$
which are linear in $S_{ij}$, one may describe the related
interaction by defining appropriate background \emph{gauge}
fields associated with the group SO(n); but in our work, we
will not follow this line.) We are here particularly keen
on the ground state degeneracy, and to facilitate
our discussion it will be assumed that the function
${\cal Q}({\bf x, S})$ in (\ref{H}) has been chosen so that
the lowest eigenvalue of $\cal H$ be equal to zero.
Now our problem is : what kind of structures (and behaviors)
for the functions ${\cal P}_i({\bf x,S})$ and ${\cal Q}({\bf x, S})$
should be assumed to make our Hamiltonian to have a
large, even infinite, zero-energy eigenspace?

In the 2d case, Aharonov and Casher \cite{Aharonov}
made an interesting observation. If one chooses
\begin{eqnarray}\nonumber
{\cal P}_i  &=& - \frac{2e}{c}A_i ({\bf x} ),
~~(i=1,2;\ {\bf x} = (x,y)) \\
{\cal Q} &=& \Big( \frac{i\hbar e}{c}
\nabla\cdot {\bf A} +\frac{e^2}{c^2} {\bf A}  
\cdot{\bf A}  \Big) -\frac{\hbar e }{c}\sigma_3 B
\end{eqnarray}
in (\ref{H}), with  arbitrary vector potential  ${\bf A}({\bf r})$  and its magnetic field $B=\p_1 A_2-\p_2 A_1$, one is led to
the Pauli Hamiltonian  
\begin{equation}\label{square}
{\cal H}   = \frac{1}{2M} H_{D}^2
\end{equation}
with
\begin{equation}\label{sigma}
H_D =   {\boldsymbol \sigma}\cdot \big(  {\bf p}-\frac{e}{c}{\bf A}\big)\ .
\end{equation}
The $\sigma$'s are usual $2\times2$ Pauli matrices.
Then, using the property of the `Dirac Hamiltonian' $H_D$,
they showed that ground state wave functions of $\cal H$
-- the states with zero energy -- can always be found
analytically, with the degree of degeneracy determined
by the total magnetic flux in accordance with the index
theorem \cite{Atiyah,Callias}. Hence, if there exists
a magnetic field approaching a nonzero constant value
asymptotically, infinitely degenerate ground states follow
always.

In
this work  we will push this idea to higher dimensional
setting to get a unified understanding on the problem
posited above (in the Euclidean space only). It is found
that there exists a special family of Hamiltonians that
enjoys the explicit ground-state integrability and encompasses
all previously known models with the same property
as special cases. New 3d models with interesting ground-state structure emerge along the way. In our 3d model with a uniform magnetic field, it is shown that entire 2d Landau-level wave functions,
separated along the field direction, make up degenerate ground states.

The rest of this work is organized as follows. In Sec. 2, a rather general discussion is offered on the structure of nonrelativistic Hamiltonians which can exhibit large ground-state degeneracy.
In Sec. 3 we consider two classes of 3d Hamiltonians for which the full space of degenerate ground states can be exhibited explicitly as in the case of the 2d Pauli Hamiltonian. One is without any magnetic field and the other is with a magnetic field. In Sec. 4 we conclude with some remarks. In Appendix, we summarize the ideas in 
arbitrary spatial dimension.

\section{Hamiltonians with large ground-state degeneracy}

Our general strategy is as follows: large ground-state degeneracy results if a given nonrelativistic Hamiltonian $\cal H$ can be
written in a square form (\ref{square}), with the related
Dirac Hamiltonian $H_D$ having a structure that admits
a large number of zero-energy eigenstates as required
by the index analysis  in open Euclidean spaces \cite{Callias}.
Here we restrict to hermitian   $H_D$ so that the spectrum  of the nonrelativistic Hamiltonian ${\cal H}$ is nonnegative.

Being interested ultimately in nonrelativistic theory, we
will write our Dirac Hamiltonian in the form
\begin{equation}\label{dirac0}
H_D =   {\boldsymbol \alpha}\cdot {\bf p}  + K({\bf x}, \boldsymbol{\alpha}, \beta)\ ,
\end{equation}
by  using hermitian Dirac matrices
${\boldsymbol \alpha} =( \gamma^0\gamma^1, \gamma^0\gamma^2,\cdots \gamma^0\gamma^n)$ 
and $\beta \equiv \gamma^0$ which satisfy the relations
$\{\alpha^i, \alpha^j\}=2\delta^{ij}\mathrm{I}$ and
$\{\beta, \alpha^i\}=0$ (with the matrices $\gamma^{\mu}$
($\mu = 0,1,\cdots, n$) satisfying the Dirac-Clifford
algebra relation $\{\gamma^\mu, \gamma^\nu\}=
-2\eta^{\mu\nu}$, $\eta^{\mu\nu}=$ diag($-1,1,1,\cdots$)).
The matrix $\beta$ is taken to have the diagonal form 
\begin{equation}
\beta = \left(\begin{array}{cc}
\mathbf{1} &0 \\ 0 &-\mathbf{1}\end{array}
\right)\ ,
\end{equation}
and the function $K({\bf x}, \boldsymbol{\alpha}, \beta)$ in
(\ref{dirac0}), a Dirac matrix polynomial, may be constrained
by the condition $\{\beta, K({\bf x}, \boldsymbol{\alpha}, \beta)\}=0$
so that we have $\{\beta, H_D\}=0$ and $[\beta, {\cal H}]=0$.
As $\beta$ is diagonal, the Hamiltonian (\ref{square}) can be written as
\begin{equation}\label{pmh}
{\cal H}  = \frac{1}{2M} H_D^2 = \left(\begin{array}{cc}{\cal H}_{+}  & 0 \\
0 & {\cal H}_{-} 
\end{array}\right)\   .
\end{equation}

With our choice of the Dirac Hamiltonian (\ref{dirac0}), eigenstates of the nonrelativistic Hamiltonian ${\cal H} $ of (\ref{square}) 
can always be chosen to have a definite $\beta$-parity, 
$  \eta = \pm 1$. For  a given energy eigenstate $\Psi$ with
the properties $\beta\Psi = \eta \Psi$ and
${\cal H} \Psi = E\Psi$ where $E>0$,
another state $\tilde\Psi \equiv H_D\Psi$ will
satisfy
\begin{equation}
\beta\tilde\Psi = -\eta\tilde \Psi\ , ~~~{\cal H} 
\tilde\Psi = \frac{1}{2M} \left(H_D \right)^3 \Psi
=E\tilde\Psi\ .
\end{equation}
Hence all nonzero energy eigenstates of ${\cal H}  $
appear pairwise, one with $\eta=+1$ and the other with $\eta=-1$ ;
this means that, when we leave out zero energy states, the two
nonrelativistic Hamiltonians given in (\ref{pmh}) are isospectral.
For zero energy states, which must be the ground states,
if exist, the situation can be different.

 Subsequent developments
depend on the spatial dimension, and we shall   study
2d problem in this section and  3d problems in the next section.  The case of higher-dimensional
spaces will be considered in the context of specific models in the Appendix.

In 2d case, Dirac matrices are assumed by
$\alpha^i = \sigma_i$ ($i=1,2$) and $\beta = \sigma_3$.
Our Dirac Hamiltonian will then have the general form
\begin{equation}\label{general}
H_D^{(2)} = \sum_{i=1}^{2}\left\{
\sigma_ip_i +\sigma_iU_i({\bf x})+i\sigma_i\beta V_i({\bf x})\right\}\ .
\end{equation}
But we here have a 2d-specific identity
$\sigma_i \beta = -i\epsilon_{ij}\sigma_j$, and so the
$U_i$ and $V_i$ terms in (\ref{general}) are not independent.
The form (\ref{sigma}), used to define the 2d Pauli Hamiltonian,
is essentially the unique possibility ; but, one may express this
by an alternative form
\begin{equation}\label{alternative}
H_D^{(2)} = \sum_{i=1}^2 \sigma_i
\left(p_i - \frac{ie}{c} \beta \epsilon_{ij}A_j({\bf x})\right)\ .
\end{equation}
In the gauge $\nabla \cdot {\bf A}=0$, we can find a scalar `potential' $\phi({\bf x})$ such that
\begin{equation} A_1 = -\frac{\hbar c }{e}\partial_2\phi, \ \ A_2=\frac{\hbar c  }{e} \partial_1 \phi \end{equation}
and so the 2d magnetic field $B=\frac{\hbar c}{e}\nabla^2\phi $. The alternative form ({\ref{alternative}) becomes
\begin{equation}\label{alternative1}
H_D^{(2)} =   -i\hbar \sum_{i=1}^2 \sigma_i
\big[ \partial_i +     \beta \partial_i \phi({\bf x})\big]\ .
\end{equation}
If we here take a
background of the form
$\phi = \frac{e}{4\hbar c}  B_0(x^2+y^2)$ as appropriate
for a uniform magnetic field $B({\bf x})= B_0$,
the above Hamiltonian $H_D^{(2)}$
reduces to the 2D version of a so-called Dirac
oscillator Hamiltonian \cite{Moshinsky} (but without
mass term). (This was noticed also in Ref. \cite{Li-Intriligator}).
 If $\Psi_0$ corresponds to a zero energy  state of
${\cal H}^{(2)} = (H_D^{(2)})^2/2M$, it should satisfy the equation

\begin{equation}\label{greq0}
H_D^{(2)}\Psi_0 = -i\hbar \sum_{i=1}^2 \sigma_i \cdot\left( \partial_i +  \beta \partial_i \phi\right)\Psi_0=0\ .
\end{equation}
With the $\beta$ parity chosen such that  $\beta\Psi_0=\eta\Psi_0$ ($\eta=\pm 1$), we may set
\begin{equation}\label{greq1}
\Psi_0(\vec x) = e^{-\eta \phi({\bf x})} {\bf F}_{\eta}({\bf x})
\end{equation}
to recast  (\ref{greq0}) into the following equation on   
the   2-component spinor ${\bf F}_\eta$:
\begin{equation} \label{greq2} \Big(\sum_{i=1}^2 \sigma_i
\partial_i \Big) {\bf F}_\eta = 0\ .  \end{equation}
If we here write 
\begin{equation}\label{greq3}
 {\bf F}_{\eta=+1}=\left(\begin{array}{c} f_{+1}\\0\end{array}\right), \ \ {\bf F}_{\eta=-1}=\left(\begin{array}{c}0\\ f_{-1}\end{array}\right)\ , 
 \end{equation}
we then see from (\ref{greq2}) that the scalar functions $f_\eta({\bf x})$ should satisfy 
\begin{equation}\label{greq4}
\left(\partial_x +i\eta\partial_y\right)f_\eta(x,y)=0\ ,
\end{equation}
that is, $f_\eta(x,y)$ should be a function of the variable
${\zeta}=x+i\eta y$ only. This was noted already in
Ref. \cite{Aharonov}, and one may use this result
together with (\ref{greq1}) to produce ground state wave
functions completely. 

The number of  independent ground states is decided by the asymptotic behavior of the scalar $\phi$ which is in turn dictated by total magnetic flux on the 2d plane.  If the magnetic field is   asymptotically
uniform, i.e.,
$B({\bf x})\rightarrow  B_0 (>0)$ as
$r=\sqrt{x^2+y^2}\rightarrow \infty$,    $\phi({\bf x})\rightarrow \frac{e}{4\hbar c}B_0 r^2$ asymptotically in the rotationally symmetric gauge. In this case, one
finds
infinitely degenerate ground states (all with the eigenvalue
$\eta =+1$), and the explicit ground state wave functions take the form
\begin{equation}\label{greq5}
\left\{\Psi_{0(n)}({\bf x}) = e^{- \phi(\bf x)}
(x+iy)^n ~;~ n=0,1,2,\cdots\right\}\ .
\end{equation}
Note that all wave functions in (\ref{greq5}) remain
normalizable as long as $(\ln r)^{-1}| \phi(\bf x)|\rightarrow \infty$ ;
with $|  \phi({\bf x})|\rightarrow c\ln r~(c>0)$, only a finite
number of zero-energy modes are allowed.  In the asymptotically uniform magnetic field, one  could have chosen the asymptotically Landau gauge potential with
$\phi({\bf x})\rightarrow \frac{e}{2 \hbar c} B_0y^2$ as
$r\rightarrow \infty$. Then, the  ground state wave function would take
the form of
continuum states
\begin{equation}
\label{greq6}
\left\{\Psi_{0(k)}({\bf x})=e^{- \phi({\bf x})}
e^{ik(x+iy)}~;~ k ~\mathrm{real}\right\} \ .
\end{equation}

\section{Models based on 3d Dirac Hamiltonians}

In 3d case, Dirac matrices $\alpha^i$ ($i=1,2,3$) and $\beta$
are provided by following $4\times 4$ matrices
\begin{equation}\label{dirac}
\alpha^1 = \left(\begin{array}{cc} 0& \sigma_1 \\ \sigma_1 & 0\end{array}\right)\ ,~~
\alpha^2 = \left(\begin{array}{cc} 0& \sigma_2 \\ \sigma_2 & 0\end{array}\right)\ ,~~
\alpha^3 = \left(\begin{array}{cc} 0& \sigma_3 \\ \sigma_3 & 0\end{array}\right)\ ,~~
\beta = \left(\begin{array}{cc} {\bf 1}& 0 \\ 0& -{\bf 1}\end{array}\right)\ .
\end{equation}
Now our hermitian Dirac Hamiltonian $H_D$ satisfying the condition  $\{\beta, H_D\}=0$ may have the form
\begin{equation}\label{3}
H_D^{(3)}=   \boldsymbol{\alpha}\cdot \Big[ {\bf p} +{\bf U}  + i  \beta {\bf V}\Big]  +\gamma^5 W({\bf x})
+i\gamma^5\beta X({\bf x})\ ,
\end{equation}
where
\begin{equation}
\gamma^5 =   -i \alpha^1\alpha^2\alpha^3   =
\left(\begin{array}{cc} 0 &  {\bf 1}\\  {\bf 1} & 0\end{array}\right)\ ,
\end{equation}
and all terms exhibited above are independent. Notice that our Dirac operator in (\ref{3}) is entirely
off-diagonal.
Here the
situation differs from the 2D case as Dirac spinors
are defined in a 4-column space in contrast to
nonrelativistic spinors defined in a 2-column space. 
Because of this,
we are led to consider a pair of nonrelativistic
Hamiltonians ${\cal H}_{\pm}^{(3)}$ (defined in
2-column spaces), according to
\begin{equation}\label{3dpm}
{\cal H}^{(3)} = \left(\begin{array}{cc}{\cal H}_{+}^{(3)} & 0 \\
0 & {\cal H}_{-}^{(3)} 
\end{array}\right)= \frac{1}{2M} \left(H_D^{(3)}\right)^2\ .
\end{equation}
Note that the subscripts $\pm$  denote the parity under 3d $\beta$. 
 In the form (\ref{3dpm}) we have natural
candidate 3d Hamiltonians which can result in large
ground state degeneracy with suitably chosen
background potentials $U_i({\bf x})$, $V_i({\bf x})$,
$W({\bf x})$ and $V({\bf x})$. Various potentials here
may be considered in connection with their specific
effects: $U_i({\bf x})$ for standard magnetic vector
potentials, and $V_i({\bf x})$ for spin-orbit-coupling-like
terms (see below), etc.

In quantum-field-theoretic investigations the Dirac
Hamiltonians similar to our form (\ref{3}) have been
discussed previously \cite{Weinberg,Niemi,Hirayama};
there, potentials are typically those related to localized
solitonic backgrounds (e.g., magnetic monopoles)
and as such they usually involve internal space
generators also. They also appear in some condensed matter literature discussing topological defects in insulators and superconductors; see Ref.\cite{TI} for instance. Zero modes of those Hamiltonians
have been studied, together with responsible
topological invariants. But, with the backgrounds of
less restricted asymptotic behaviors (e.g. with unbounded potentials),  and especially
with an eye on their physical significance in
nonrelativistic Hamiltonian contexts, the analysis
is not so simple and up to our knowledge no
systematic study has been made. Hence, leaving
such to our future study, we shall below concentrate
on some special class of Hamiltonians, which enjoy
explicit ground-state integrability and so can be
used to exhibit some of the expected features.
As we shall see, our model Hamiltonians, which have
not been seriously considered in a relativistic setting,
turn  out to have some interesting nonrelativistic
contents.

Our first model in 3d is obtained from (\ref{3}) by keeping
only the potential  ${\bf V}({\bf x})$ in the form
${\bf V}({\bf x})=-\hbar \nabla \phi({\bf x})$,
$\phi({\bf x})$ being arbitrary. That is, we consider
the Dirac Hamiltonian
\begin{equation}\label{3ddrac}
H_D^{(3)} = \boldsymbol{\alpha}\cdot\big[ 
{\bf p}-i \hbar \beta \nabla\phi({\bf x})\big]\ ,
\end{equation}
and this model may in fact be considered in
arbitrary spatial dimensions.  
In 3d case, a simple calculation using the Dirac
matrices in (\ref{dirac}) yields   the 3d Hamiltonian (\ref{3dpm})    which   consists of a pair of nonrelativistic Hamiltonians
\begin{equation}\label{18}
{\cal H}_{\pm}^{(3)} =\frac{1}{2M}\Big({\bf p}^2 \mp  2\hbar
(\nabla\phi\times{\bf p})\cdot \boldsymbol{\sigma}
+  \hbar^2\nabla\phi\cdot  \nabla\phi\mp
\hbar^2 \nabla^2\phi\Big)\ .
\end{equation}
If we here define the `electric' field by
${\bf E}({\bf x})=- \nabla\phi({\bf x})$, the second term
in the right hand side of (\ref{18}) obviously describes a
spin-orbit coupling. Some recently discussed models with
three-dimensional Landau-level-like structures correspond
to special cases of this model, i.e.,
$\phi({\bf x})=(\mathrm{const.})(x^2+y^2+z^2)$ in  Refs. \cite{Li-Wu,Li-Intriligator}, and
$\phi({\bf x})=(\mathrm{const.})z^2$ in the model of
Ref. \cite{Li-Zhang}. In our discussion, however, the
detailed profile of the potential $\phi({\bf x})$ will be
left largely arbitrary. As far as ground-state degeneracy
structure is concerned, what matters is the asymptotic
behavior of the potential -- in our framework, a particular
symmetry in the background potential is irrelevant.
But our approach cannot say anything as regards
possible excited-state degeneracy.
 
The zero-energy eigenfunctions   $\Psi_0$ of  the Dirac Hamiltonian (\ref{3ddrac}), as needed for the ground states of the Hamiltonian (\ref{18}), can be written as      
\begin{equation}\label{3dgreq1}
\Psi_0(\vec x) = e^{-\eta \phi({\bf x})} {\bf F}_{\eta}({\bf x})
\end{equation}
with the $\beta$ parity $\eta$. Then the  
 wave function ${\bf F}_\eta({\bf x})$ satisfies
\begin{equation}\label{3deq1}
\boldsymbol{\alpha} \cdot  \nabla {\bf F}_{\eta}({\bf x})=0\ .
\end{equation}
For a given large-$|{\bf x}|$ behavior of $\phi({\bf x})$, we usually
obtain a physically acceptable wave function from (\ref{3deq1}) only
with a particular sign for $\eta$. Aside from this, finding ground
state wave function is now down to solving the background-free
equation (\ref{3deq1}).
In this case, we  again write
${\bf F}_{\eta = +1}=\left(\begin{array}{c}{\bf f}_+ \\0\end{array}\right)$
and 
${\bf F}_{\eta = -1}=\left(\begin{array}{c}0\\{\bf f}_- \end{array}\right)$,
but this time ${\bf f}_\eta $ make 2-component spinors.
Then the Dirac equation (\ref{3deq1}), with the Dirac matrices
in (\ref{dirac}) used, reduces to the 2-component spinor
equation
\begin{equation}\label{3deq2}
\boldsymbol{\sigma} \cdot  \nabla {\bf f}_\eta =0  
 \ .
\end{equation}
Since 
\begin{equation}\label{3deq3}
\left(\boldsymbol{\sigma}\cdot \nabla\right)^2 = 
 \partial^2_x +
 \partial^2_y +
 \partial^2_z 
\equiv \ \nabla^2\ ,
\end{equation}
the 2-component spinor ${\bf f}$ should have harmonic functions as its components.
Now, in view of the identity
\begin{equation}
\nabla^2(z+ix\cos u+iy\sin u)^p=0\ ,
~(u~ \mathrm{real})
\end{equation}
one may write the general solution to (\ref{3deq2}) in the form
\cite{Whittaker}
\begin{equation}\label{29}
{\bf f}({\bf x}) = \int_{-\pi}^{\pi} h(z+ix\cos u+iy\sin u,u)
{\boldsymbol\chi}(u) \mathrm{d} u\ .
\end{equation}
Here $h$ is an arbitrary function,  and  ${\boldsymbol\chi}(u)$ represents a 2-spinor satisfying
\begin{equation}\label{30}
(\sigma_3+i\sigma_1\cos u +i\sigma_2\sin u){\boldsymbol\chi}(u)=0\ ,
\end{equation}
whose solution  has the following form
\begin{equation}\label{32}
{\boldsymbol\chi}(u)=\left(\begin{array}{c}1\\ie^{iu}\end{array}\right)\ .
\end{equation}

In (\ref{29}) any function $h$ is allowed as long as, when
used in (\ref{3dgreq1}), it gives rise to a physically acceptable
ground-state wave function $\Psi_0({\bf x})$\ . Given
the potential with the asymptotic behavior
$\phi({\bf x})\rightarrow cr^2~(c>0)$ as
$r=\sqrt{x^2+y^2+z^2}\rightarrow \infty$
(as in the model of Refs. \cite{Li-Wu,Li-Intriligator}),
we must choose $\eta=+1$ and then take for $h$
following polynomial forms
\begin{equation}
h=(z+ix\cos u +iy\sin u)^l e^{imu}\ ,
(l=0,1,2,\cdots ; m=-l,-l+1,\cdots l)\ .
\end{equation}
Then, using (\ref{32}) and performing the integration over
$u$, we obtain the results for ${\bf f}({\bf x})$ written using
spherical harmonics :
\begin{equation}\label{34}
{\bf f}_{+1}({\bf x})=\left(\begin{array}{c}
\sqrt{\frac{l+m+1}{2l+1}}~r^lY_{lm}(\theta,\phi)\\
\sqrt{\frac{l-m}{2l+1}}~r^lY_{l,m+1}(\theta,\phi)
\end{array}\right)
\end{equation}
We thus have infinitely degenerate ground states in any
such background. Actually all forms in (\ref{34}) leads
to normalizable wave functions as long as the asymptotics
of the background potential is such that 
$(\ln r)^{-1}|q\phi(\bf x)|\rightarrow \infty$ as $r\rightarrow \infty$,
and, if $|q\phi({\bf x})|\rightarrow c\ln r~(c>0)$ as 
$r \rightarrow \infty$, only a finite number of zero-energy
modes survive. The detailed form of the background
potential at finite $r$ is not important for our discussion.

On the other hand, with the background
potential having the asymptotic behavior
$\phi({\bf x}) \rightarrow cz^2 ~(c>0)$ as $r\rightarrow \infty$
(as in the model considered in Ref. \cite{Li-Zhang}),
we obtain bounded ground-state wave functions with
$f_+({\bf x})$ given by the continuum
\begin{equation}\label{35}
{\bf f}_{+1}({\bf x}) = e^{k(z+i x\cos u + i y \sin u)}
\left(\begin{array}{c}1\\ie^{iu}\end{array} \right)\ ,
~(k,u~\mathrm{real})\ .
\end{equation}
Note that, integrating this form over $u$ with weight $e^{imu}$, one may also take for ${\bf f}_{+1}$ the following expression:
\begin{equation} {\bf f}_{+1}({\bf x}) = e^{kz+ im\varphi} \left(
\begin{array}{c} J_m(k\rho) \\ iJ_{m+1}(k\rho)\end{array}\right) , \ (k \ {\rm real} \ ;  m=0,\pm 1, \pm 2 , \cdots )\end{equation}
where   $x+iy = \rho e^{i\varphi}$ and $J_m(k\rho)$ denotes Bessel functions.

 For further discussions on these wave functions
including possible physical applications, see Ref.
\cite{Li-Wu,Li-Intriligator,Li-Zhang}. In 3d, the two expressions we have chosen above for the asymptotic form of  the scalar $\phi$ are not gauge equivalent as the scalar field is not connected to the gauge field in 3d and so the physics for these two cases are different. 

Let us now consider a little more complicated 3d Dirac Hamiltonian which has both gauge field and the scalar gradient as in 
\begin{equation} \label{3ddirac1} H_D = \boldsymbol{\alpha}\cdot \Big(\big[ {\bf p} - \frac{e}{c} {\bf A}({\bf x})\big] - i\hbar \beta \nabla \phi({\bf x)}   \Big) 
\end{equation}
Under this choice, the related nonrelativistic Hamiltonians (\ref{3dpm}) read as
\begin{equation} {\cal H}_\pm = \frac{1}{2M}
\Big( [{\bf p}-\frac{e}{c}{\bf A}]^2 \mp2\hbar \nabla\phi\times
[{\bf p}-\frac{e}{c}{\bf A}]\cdot\boldsymbol{\sigma} -\frac{\hbar e}{c} \boldsymbol{\sigma}\cdot {\bf B} + \hbar^2 \nabla\phi\cdot\nabla\phi\mp \hbar^2\nabla^2\phi\Big)\ .
\end{equation}
where ${\bf B}=\nabla\times {\bf A}$. 
If $\Psi_0$ corresponds to a zero-energy ground state of
${\cal H} $, it should satisfy the equation
\begin{equation}\label{3ddgreq0}
H_D\Psi_0 = \boldsymbol{\alpha}\cdot\left(\big[-i\hbar\nabla
- \frac{e}{c} {\bf A}({\bf x})\big]- i\hbar \beta \nabla\phi\right)\Psi_0=0\ .
\end{equation}
With the $\beta$ parity $\eta$ chosen so that  $\beta\Psi_0=\eta\Psi_0$ ($\eta=\pm 1$), we may set
\begin{equation}\label{3ddgreq1}
\Psi_0(\vec x) = e^{-\eta \phi({\bf x})} {\bf F}_{\eta}({\bf x})
\end{equation}
to recast  (\ref{3ddgreq0}) into the following equation on ${\bf F}_{\eta}({\bf x})$:
\begin{equation}\label{3ddgreq2}
\boldsymbol{\alpha} \cdot \big[-i\hbar\nabla
- \frac{e}{c} {\bf A}({\bf x})\big]  {\bf F}_{\eta}({\bf x})=0\ .
\end{equation}
For a given large-$|{\bf x}|$ behavior of $\phi({\bf x})$, we usually
obtain a physically acceptable wave function from (\ref{3ddgreq1}) only
with a particular sign for $\eta$. Aside from this, finding ground
state wave function is now down to solving the simpler
equation (\ref{3ddgreq2}). Writing 
$\label{3d} {\bf F}_{+1}=  \left(\begin{array}{c}   {\bf f}_{+1}  \\  0 \end{array}\right)   $ and $ {\bf F}_{-1}=    \left(\begin{array}{c} 0 \\   {\bf f}_{-1}\end{array}\right) $,
(\ref{3ddgreq2}) reduces to the 2-component spinor equations
\begin{equation} \label{3dlandau} \boldsymbol{\sigma}\cdot (  \nabla -\frac{i e }{\hbar c}{\bf A}) {\bf f}_\eta = 0 \end{equation}
For some magnetic flux, one could find normalizable zero-energy ground states if the scalar function $\phi$ and the $\beta$ parity $\eta$ were   chosen suitably.  

Since a direct analysis of (\ref{3dlandau}) with an arbitrary 3d vector potential ${\bf A}({\bf x})$ is impossible, let us restrict our attention to the case of a strictly 2d vector potential
\begin{equation} \label{3dvec}
{\bf A} = (A_1(x,y), A_2(x,y), A_3=0) 
\end{equation}
and rewrite (\ref{3dlandau}) as
\begin{equation} \label{3dlandau1}
 (\tilde H_D^{(2)} + i   p_3  ){\bf f}(x,y,z)= 0 , \ \ \tilde H_D^{(2)} = i\sigma_3 \sum_{i=1}^2 \sigma_i (p_i -\frac{e}{c} A_i) 
\end{equation}

As $[\tilde H^{(2)}_D,p_3 ]=0$, we choose the eigenfunctions of $p_3$ as the solution of (\ref{3dlandau1}). Also note that   $\tilde H_D^{(2)}$ is another hermitian Dirac Hamiltonian related to a 2d Pauli Hamiltonian, and so all normalizable energy engenfunctions of this operator can be found in principle. For the solutions of (\ref{3dlandau1}) with $(p_3)'=0$ we need   zero-energy eigenfunctions of $\tilde H^{(2)}_D$: they were found in Sec.2. But, for the solutions with $(p_3)'\neq 0$, we here need also nonzero-energy eigenfunctions of $\tilde H^{(2)}_D$. As nonzero eigenvalues of this Dirac operator should appear pairwise with opposite signs, we may express the related eigenvalue equations schematically as 
\begin{equation}  \tilde H^{(2)}_D   {\bf h}_\ell = k_\ell  {\bf h}_\ell\ , \  \ \tilde H^{(2)}_D   \s_3 {\bf h}_\ell = - k_\ell  \s_3 {\bf h}_\ell \ .  \end{equation}
where $\ell$ labels all independent nonzero-energy engenstates. Then we can represent the $(p_3)'=\pm i k_\ell (\neq 0)$ solutions to (\ref{3dlandau1}) by the form
\begin{equation} {\bf f}= e^{ -\frac{ k_\ell}{\hbar } z}{\bf h}_\ell (x,y), \ \  {\bf f}= e^{ +\frac{ k_\ell}{\hbar } z} \sigma_3{\bf h}_\ell(x,y) , \end{equation}
Note that  wave functions with imaginary eigenvalue of $p_3=-i\hbar \partial_3$ would be out of consideration usually. However, in our case,  there is additional factor $e^{-\eta \phi}$ entering the wave function (\ref{3ddgreq1}), and an appropriate choice of $\eta$ and $\phi$ (imagine the scalar $\phi$ having the asymptotic behavior $\phi\sim cz^2$) would provide sufficient fall off for finite  $k_\ell$ at large $|z|$, making the 
ground wave function $\Psi_0$ in ({\ref{3ddgreq1})   normalizable.

Especially, if the vector potential (\ref{3dvec}) is chosen to be that of the   uniform magnetic field ${\bf B}= B\hat{z}$,  we can represent 
the eigenstates of $\tilde H^{(2)}_D$ in terms of the well-known Landau-level wave functions. Indeed, we then have $\tilde H^{(2)}_D$ (in the symmetric gauge) expressed as  
\begin{equation} \tilde H^{(2)}_D= -\sqrt{\frac{2\hbar eB}{c} }\left( \begin{array}{cc} 
0 & \bar a  \\ a & 0 \end{array}\right)
\end{equation}
where $a, \bar a $ denote  the creation and annihilation operators 
\begin{equation}
a =  \sqrt{\frac{2\hbar c}{eB}}  ( \partial_{\bar\zeta}+\partial_{\bar\zeta}\chi ) ,\ \  \bar a = - \sqrt{\frac{2\hbar c}{eB} }( \partial_{ \zeta}-\partial_{ \zeta}\chi ) , 
\end{equation}
with $\zeta=x+iy, A_i = -\frac{\hbar c}{e} \e_{ij}\partial_j \chi$, and  $\chi=\frac{eB}{4\hbar c}\zeta\bar\zeta$. The eigenstates and eigenvalues of $\tilde H^{(2)}_D$ are, respectively,
\begin{equation} \label{3dlandaulevel} {\bf h}_{\ell,n} (x,y) = \left(\begin{array}{c} \bar a^{|\ell|} \Phi_n \\
-\ell \bar a^{|\ell|-1} \Phi_n \end{array}\right), \ \ k_\ell = \ell\sqrt{\frac{2\hbar eB}{  c}}  
\end{equation}
where $\ell=0, \pm 1, \pm 2, \cdots$, and    $\Phi_n= \zeta^n e^{-\chi}$ with $ n=0,1,2,\cdots $. 
Especially, a choice $\phi= cz^2$ would make all 2d Landau levels be available for ground states of the 3d problem; but, as they are to be multiplied by the factor $e^{\pm \frac{k_\ell}{\hbar} z - cz^2}$, the excited 2d Landau levels appear separated along the z-axis. As such degenerate stacking of all 2d landau levels is possible, ground state degeneracy of this 3d system is infinite times larger than that of the 2d Landau system. The 2d Landau level functions in (\ref{3dlandaulevel}) can also be used to find the ground state wave functions when the scalar field $\phi$ has somewhat different asymptotic behaviors. In the case that we have $\phi\sim \gamma|z| (\gamma>0)$ asymptotically, only a partial set of the above 2d Landau levels, that is, those with the Landau level index $\ell$ satisfying the condition $|k_\ell/\hbar|<\gamma$, would be acceptable for the ground states of the 3d system.

\section{Concluding Remarks}

In this paper we have shown that a fruitful way to study
quantum systems exhibiting large ground-state degeneracy
is to look for a connection to a Dirac operator. We are
then led to a particular class of Hamiltonians, which
exhibit explicit ground-state integrability (for any background
potential with the given asymptotic behavior) and at
the same time serve as a unified framework for some of the
recently proposed Hamiltonians in condensed matter
physics. Needless to say, if the  models considered in this work turn out to have some direct experimental relevance, it will be most welcome.  To gain further insight, it should be desirable to have our analysis extended to the 3d model based on the full 3d Dirac Hamiltonian (\ref{3}). Considering our model in a curved space will be
another interesting future problem.

\vskip 1cm
\centerline{\bf\large Acknowledgments}
\vskip 5mm
The work of CL was partly performed while he was visiting
the Korea Institute for Advanced Study for the year
2012/2013, and he is grateful to Prof. D. Kim for making
this visit possible.   This work is supported by the National Research Foundation of Korea Grants   2006-0093850 (KL), 2009-0084601 (KL), and 2005-0049409 (KL) through the Center for Quantum Space-Time(CQUeST) of Sogang University.

\appendix
\numberwithin{equation}{section}

\section{Arbitrary Dimension}

We shall comment on the case with a
higher-dimensional-space version of the  
Hamiltonian ${\cal H}$ obtained from the  $n$-dimensional generalization of the Dirac Hamiltonian (\ref{3ddrac}). For this, following Ref. \cite{Li-Intriligator}, it is
convenient to introduce rank-$k$ $\Gamma$-matrices,
$\Gamma_i^{(k)}$, $i=1,2,\cdots,2k+1$ which are $2^k$ by
$2^k$ and satisfy $\{\Gamma_i, \Gamma_j\}=2\delta_{ij}$ ;
these matrices can be constructed iteratively using the
recursive formulas
\begin{equation}
\Gamma_i^{(k)} = \left(\begin{array}{cc}0&\Gamma_i^{(k-1)}\\
\Gamma_i^{(k-1)} &0\end{array}\right)
~(i\le k-1)\ ,~~
\Gamma_{2k}^{(k)}=\left(\begin{array}{cc}
0&-i\mathrm{I}\\i\mathrm{I}&0\end{array}\right)\ ,
~~\Gamma_{2k+1}^{(k)}=\left(\begin{array}{cc}
\mathrm{I}&0\\0&-\mathrm{I}\end{array}\right)\ ,
\end{equation}
starting from rank-$1$ $\Gamma$-matrices
$\Gamma_i^{(1)}=\sigma_i$ ($i=1,2,3$).   
Using these $\Gamma$-matrices, the SO(d) generators
in the fundamental spinor representation can be given.
In $d=2k+1$-dimensional space, one can take
\begin{equation}
S_{ij}=-S_{ji} = -\frac{i}{4}\left[\Gamma_i^{(k)}, \Gamma_j^{(k)}
\right]~~(1\leq i<j\leq 2k+1)
\end{equation}
for the rotation generators. On the other hand, in
$d=2k$-dimensional space, we have two inequivalent set
of SO(d) generators, $\{S_{ij}\}$ and $\{S'_{ij}\}$, which
can be identified with
\begin{eqnarray}
\mathrm{Set~1~:~} S_{ij}&=&-S_{ji} =\left[
\begin{array}{c}-\frac{i}{4}[\Gamma_i^{(k-1)}, \Gamma_j^{(k-1)}
]\ , \mathrm{~for~} 1\leq i<j\leq 2k-1 \\
\frac12 \Gamma_i^{(k-1)}~,~ \mathrm{for~}j=2k
~\mathrm{and}~1\leq i\leq 2k-1\end{array}\right.\ ,\\
\mathrm{Set~2~:~} S'_{ij}&=&-S'_{ji} =\left[
\begin{array}{c}-\frac{i}{4}[\Gamma_i^{(k-1)}, \Gamma_j^{(k-1)}
]\ , \mathrm{~for~} 1\leq i<j\leq 2k-1 \\
-\frac12 \Gamma_i^{(k-1)}~,~ \mathrm{for~}j=2k
~\mathrm{and}~1\leq i\leq 2k-1\end{array}\right.\ .
\end{eqnarray}
We can also specify our Dirac matrices $\alpha^i$ and $\beta$
in even- or odd-dimensional space as follows :
in 2k-dimensional space, take the $2^k \times 2^k$ matrices
\begin{equation}\label{40}
\alpha^i = \Gamma_i^{(k)}~~(i=1,\cdots, 2k)\ ,
~~\beta = \Gamma^{(k)}_{2k+1}\ ,
\end{equation}
and, in $(2k+1)$-dimensional space, use the expressions
(which are $2^{k+1}$ by $2^{k+1}$)
\begin{equation}\label{41}
\alpha^i = \left(\begin{array}{cc} 0 &\Gamma_i^{(k)}
\\ \Gamma_i^{(k)} &0\end{array}\right)~~(i=1,\cdots,2k+1)\ ,
~~~\beta = \left(\begin{array}{cc}\mathrm{I}&0
\\0&-\mathrm{I}\end{array}\right)\ .
\end{equation}

Inserting these Dirac matrices in the definition  
for 
\begin{equation} H_D = \boldsymbol{\alpha}\cdot \Big[{\bf p} - i\hbar \beta \nabla \phi({\bf x})\Big] 
\end{equation}
  we come up with the   Hamiltonian
${\cal H}=\frac{1}{2M} H_D^2$ in arbitrary spatial
dimensions.  In term of two components ${\cal H}_\pm $ in  (\ref{pmh}), we have explicitly
\begin{eqnarray}\nonumber
 \underline{d=2k} & & \\\nonumber
& & 2M{\cal H}_+ =  {\bf p}^2  -4\hbar S_{ij}(\partial_i\phi)p_j
+ \hbar^2  \nabla\phi\cdot \nabla\phi 
-     \hbar^2 \nabla^2\phi \\
& & 2M{\cal H}_- =
 {\bf p}^2+4\hbar S'_{ij}(\partial_i\phi)p_j
+ \hbar^2  \nabla\phi\cdot \nabla\phi
+  \hbar^2  \nabla^2\phi , \label{2kd} \\
\underline{d=2k+1} & &   \nonumber \\
& & 2M{\cal H}_+ = {\bf p}^2-4\hbar S_{ij}(\partial_i\phi)p_j
+ \hbar^2  \nabla\phi\cdot \nabla\phi 
-  \hbar^2  \nabla^2\phi \nonumber \\
& & 2M{\cal H}_- =  {\bf p}^2+4\hbar S_{ij}(\partial_i\phi)p_j
+ \hbar^2  \nabla\phi\cdot \nabla\phi
+  \hbar^2 \nabla^2\phi \ \label{2k1d}. 
\end{eqnarray}
For $k=1$ with (\ref{2kd}), setting $S_{12}=-S'_{12}=\frac12$
produces the 2D Pauli Hamiltonian with 
$A_i = -\epsilon^{ij}\partial_j\phi$ ; taking $k=1$ in (\ref{2k1d})
and setting $S_{ij}=\frac12 \epsilon^{ijk}\sigma_k$ leads to
our earlier 3D expression.

For the ground-state wave functions one may solve   the
zero-energy Dirac equation $H_D\Psi_0=0$ by setting   $\Psi_0 = e^{ -  \beta \phi} {\bf F}$.    Then, with the definite choice of the $\beta$ parity   $\eta=\pm 1$,   we again obtain the equation (\ref{3deq1}) for ${\bf F}_\eta$.
If we here write ${\bf F}_{\eta=1}=
\left(\begin{array}{c}{\bf f}_{+1}\\0\end{array}\right)$
and ${\bf F}_{\eta=-1}=
\left(\begin{array}{c}0\\{\bf f}_{-1}\end{array}\right)$,
${\bf f}_{\pm 1}$ in $d=2k$-dimensional
($d=2k+1$-dimensional) space will have $2^{k-1}$ columns
($2^{k}$ columns) and must satisfy the following equations:
\begin{eqnarray}
\underline{d=2k}~&:&
~\left(\sum_{i=1}^{2k-1}\Gamma_i^{(k-1)}\partial_i
+i\eta\partial_{2k}\right) {\bf f}_\eta=0\ ,\\
\underline{d=2k+1}~&:&~ \left(\sum_{i=1}
^{2k+1}\Gamma_i^{(k)}\partial_i\right){\bf f}_\eta=0
~~(\mathrm{for~both~\eta=\pm1})\ .
\end{eqnarray}
Note that ${\bf f}_\eta$, in both even- and odd-dimensional
spaces, should have harmonic functions as its components.
Hence, generalizing (\ref{29}), we may express the column
function ${\bf f}_\eta$ in higher dimensional space by an
integral
\begin{equation}\label{46}
{\bf f}_\eta = \int_{{\bf u}\in S^{d-2}}
h(x_d +i{\bf u}\cdot{\bf x}) \boldsymbol{\chi}_\eta
({\bf u}) ~\mathrm{d}\Omega_{d-2}
\end{equation}
where ${\bf u}\cdot {\bf x} \equiv u^1x^1+\cdots +u^{d-1}x^{d-1}$,
$\bf u$ denoting a vector which can take values on the sphere
$S^{d-2} : (u^1)^2+\cdots +(u^{D-1})^2=1$, and
$\boldsymbol{\chi}_\eta({\bf u})$ represents a column vector
satisfying the condition
\begin{eqnarray}\label{47}
\underline{d=2k}~&:&
~\left(\sum_{i=1}^{2k-1}u^i\Gamma_i^{(k-1)}\right)
\boldsymbol{\chi}_\eta({\bf u})=-\eta\boldsymbol{\chi}_\eta({\bf u})\ ,\\\label{48}
\underline{d=2k+1}~&:&~ \left(\sum_{i=1}
^{2k}iu^i\Gamma_i^{(k)}+\Gamma_{2k+1}^{(k)}\right)
\boldsymbol{\chi}_\eta({\bf u})=0
~~(\mathrm{for~both~\eta=\pm1})
\end{eqnarray}
in even-and odd-dimensional spaces, respectively.
Note that (\ref{47}) and (\ref{48}) may be written as
eigenvector equations involving SO(D) generator matrices,
viz.,
\begin{eqnarray}
\underline{d=2k} &:& \left(\sum_{i=1}^{2k-1} u^i S_{i,2k}\right)
\boldsymbol{\chi}_{+1}({\bf u}) =-\frac12
\boldsymbol{\chi}_{+1}({\bf u})\ ,  \nonumber \\
& &  
\left(\sum_{i=1}^{2k-1} u^i S'_{i,2k}\right)
\boldsymbol{\chi}_{-1}({\bf u}) =-\frac12
\boldsymbol{\chi}_{-1}({\bf u})\ ,~~~~~\\
\underline{d=2k+1} &:& \left(\sum_{i=1}^{2k} u^i S_{i,2k+1}\right)
\boldsymbol{\chi}_{\pm1}({\bf u}) =-\frac12
\boldsymbol{\chi}_{\pm1}({\bf u})\ .
\end{eqnarray}
So one can use group theory to fix $\boldsymbol{\chi}_\eta(u)$ here~\cite{vilenkin}.
Then, depending on the asymptotic behaviors of the
background potential $\phi({\bf x})$, one may consider
polynomial or/and exponential types for the function
h in (\ref{46}), just as in 2d and 3d cases treated
earlier. This way, spinor wave functions analogous to
the form (\ref{34}) (but now involving hyper-spherical
harmonics~\cite{vilenkin}) or to the continuum expression (\ref{35})
can be obtained ; using these in (\ref{3dgreq1}) will lead to
normalizable (or at least bounded) ground-state wave
functions only if the background potential $\phi({\bf x})$
has `right' asymptotic behaviors. For some of the explicit
expressions regarding these wave functions, readers may
consult Refs. \cite{Li-Wu,Li-Intriligator,Li-Zhang}
where they are discussed in the context of specially chosen
background configurations.

\end{document}